\documentclass[aps,prl,twocolumn,amsmath,showpacs,amssymb,superscriptaddress,nofootinbib]{revtex4-1}

\usepackage{graphicx}% Include figure files
\usepackage{dcolumn}% Align table columns on decimal point
\usepackage{bm}% bold math

\usepackage{color}
\usepackage{amsfonts}
\usepackage{amssymb}
\usepackage{amsmath}

\newcommand{\gton}{\mathrel{\lower.9ex \hbox{$\stackrel{\displaystyle
>}{\sim}$}}}
\newcommand{\lton}{\mathrel{\lower.9ex \hbox{$\stackrel{\displaystyle
<}{\sim}$}}}
\newcommand{\sNN}{s_{NN}}

\newcommand{\dif}{\mathrm{d}}

\begin{document}

\title{Effects of $\rho$-meson width on pion distributions in
  heavy-ion collisions}
\author{Pasi Huovinen}
\affiliation{Institute of Theoretical Physics, University of Wroclaw,
PL-50204 Wroc\l aw, Poland}
\author{Pok Man Lo}
\affiliation{Institute of Theoretical Physics, University of Wroclaw,
PL-50204 Wroc\l aw, Poland}
\affiliation{Extreme Matter Institute EMMI, GSI, D-64291 Darmstadt, Germany}

\author{Micha\l{}  Marczenko}
\affiliation{Institute of Theoretical Physics, University of Wroclaw,
PL-50204 Wroc\l aw, Poland}

\author{Kenji Morita}
\affiliation{Yukawa Institute for Theoretical Physics, Kyoto University, 
             Kyoto 606-8502, Japan}
\affiliation{Institute of Theoretical Physics, University of Wroclaw,
PL-50204 Wroc\l aw, Poland}

\author{Krzysztof Redlich}
\affiliation{Institute of Theoretical Physics, University of Wroclaw,
PL-50204 Wroc\l aw, Poland}
\affiliation{Extreme Matter Institute EMMI, GSI, D-64291 Darmstadt, Germany}

\author{Chihiro Sasaki}
\affiliation{Institute of Theoretical Physics, University of Wroclaw,
PL-50204 Wroc\l aw, Poland}

\begin{abstract}
The influence of the finite width of $\rho$ meson on the pion momentum
distribution is studied quantitatively in the framework of the
S-matrix approach combined with a blast-wave model to describe
particle emissions from an expanding fireball. We find that the proper
treatment of resonances which accounts for their production dynamics
encoded in data for partial wave scattering amplitudes can
substantially modify spectra of daughter particles originating in
their two body decays. In particular, it results in an enhancement of
the low-$p_T$ pions from the decays of $\rho$ mesons which improves
the quantitative description of the pion spectra in heavy ion
collisions obtained by the ALICE collaboration at the LHC energy.
\end{abstract}

%\pacs{25.75.-q, 25.75.Ld, 12.38.Mh, 24.10.Nz}

\maketitle

Recent measurements of the transverse momentum, $p_T$-distributions of
identified particles produced in $\sqrt{\sNN} = 2.76$ TeV Pb+Pb
collisions at CERN Large Hadron Collider (LHC)~\cite{Abelev:2013vea}
revealed an excess of low-momentum ($p_T \lton 0.3$ GeV) pions over
the conventional fluid-dynamical
calculations~\cite{Abelev:2013vea,Begun:2013nga,Molnar:2014zha}.

It is well known that pions originating from decays of resonances
have a steeper $p_T$-distribution than the thermal
pions~\cite{Sollfrank:1991xm}, and that they provide a dominant
contribution to the spectrum at low transverse momentum. Thus, resonance
decays require a particular attention when modeling spectra of
particles originating from an expanding thermal fireball.

In fluid-dynamical calculations, the interacting hadrons are usually
described by the hadron resonance gas (HRG), where the system is
modeled as a gas of free hadrons with resonances considered as
particles with vanishing widths. This approximation yields reasonable
description of the bulk properties of the hadronic
medium~\cite{Venugopalan:1992hy,Gerber:1988tt,Weinhold:1997ig,n1}.
The HRG model also provides a very satisfactory description of
particle yields measured in heavy ion collisions~\cite{f1,f2,f3,f4, cleymans, Becattini:1997rv, Becattini:1997ii, Kisiel:2005hn, 
Chojnacki:2011hb},
as well as the hadronic equation of state and some fluctuation
observables obtained in lattice QCD
(LQCD)~\cite{Karsch:2003vd,Huovinen:2009yb,k1,k2,k3}. However, as we
show in this letter, when $p_T$-differential observables are involved,
a more refined approach may be necessary.

To properly address the dynamics of hadrons, the effect of resonance
width must be included. A conventional way is to impose a Breit-Wigner
distribution on the resonance mass. Unfortunately, this approach
proves to be too crude in many circumstances. For example, for a broad
resonance like the $\sigma$ meson~\cite{Broniowski:2015oha}, or the
(yet-to-be-confirmed) $\kappa$ meson~\cite{Friman:2015zua}, the
Breit-Wigner approach can give misleading results on the resonance
contribution to the thermodynamics.

We thus take a more fundamental approach to evaluate the properties of
interacting hadrons based on the S-matrix formulation of Dashen, Ma
and Bernstein~\cite{Dashen:1969ep}. For elastic scatterings, the
interaction part of the partition function reduces to the
Beth-Uhlenbeck form for the second virial coefficient, expressed in
terms of the scattering phase shifts~\cite{Beth:1937zz}. In the
context of heavy-ion physics, this approach has been applied to
evaluate the contribution of 
$\pi N$~\cite{Venugopalan:1992hy,Denisenko:1986bb,Weinhold:1997ig},
$\pi\pi$~\cite{Venugopalan:1992hy,Broniowski:2015oha}, and $\pi K$
interactions~\cite{Venugopalan:1992hy,Friman:2015zua} to the
thermodynamics of hadronic matter, and to analyse the resonance
production~\cite{Broniowski:2003ax}.

In this letter, to make the effects of resonance width on particle
$p_T$-spectra more tractable, we concentrate on the $\pi\pi$
system. As shown in
Refs.~\cite{Venugopalan:1992hy,Broniowski:2015oha}, the effects of the
scalar-isoscalar and the scalar-isotensor channels largely cancel each
other. This cancellation remains when the single particle distribution
of pions is evaluated. Thus for our purposes it is sufficient to
consider only the vector-isovector channel, i.e.\ the channel of the
$\rho$ meson.

%%%%%%%%%%%%%%%%%%%%% FIGURE %%%%%%%%%%%%%%%%%%%%%%%%%%%%%%%%
\begin{figure*}[t]
\begin{minipage}{0.497\textwidth}
 \hspace*{-4.5mm}
 \includegraphics[width=1.06\textwidth]{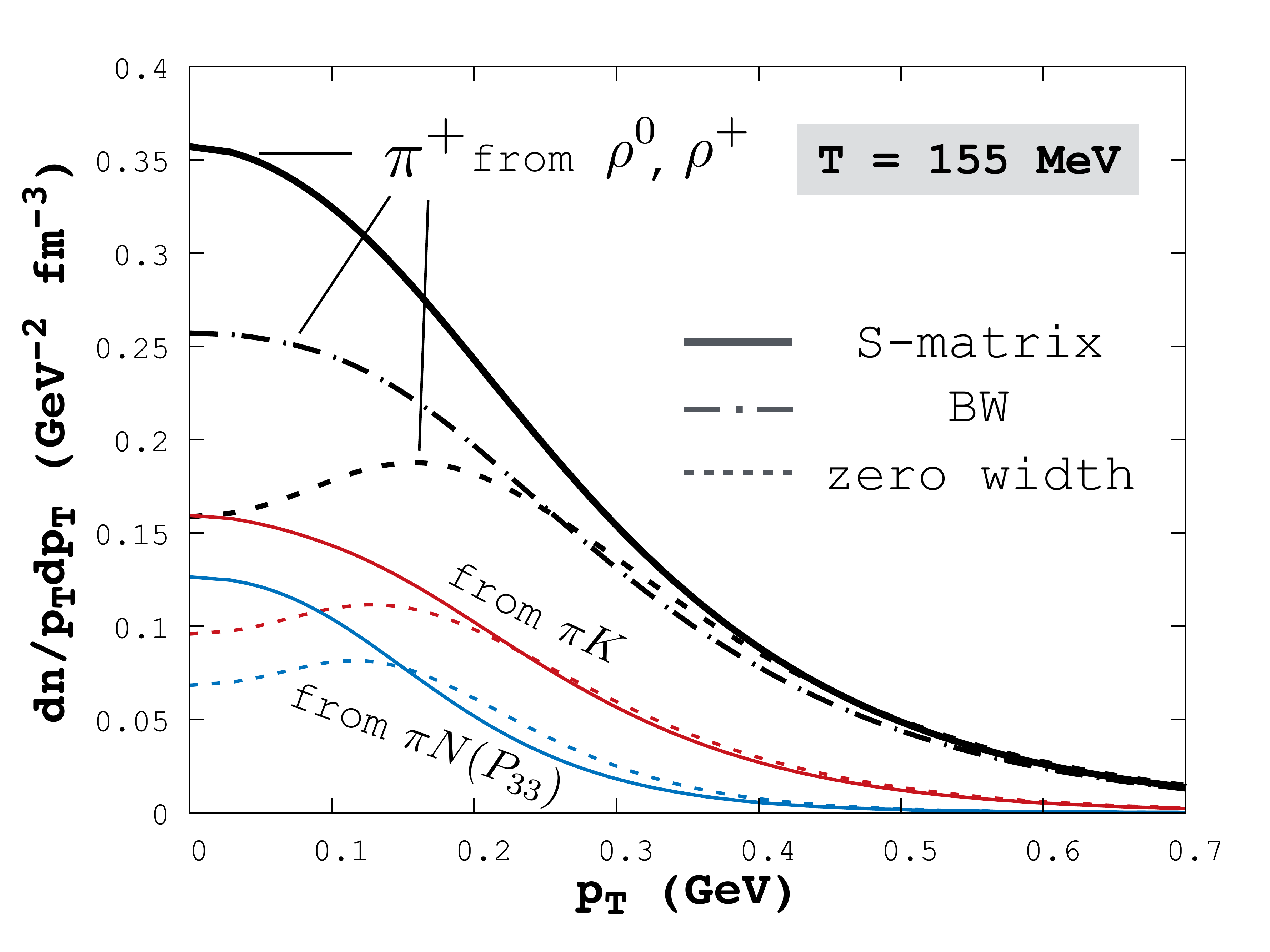}
\end{minipage}
 \hfill
\begin{minipage}{0.497\textwidth}
 \hspace*{-2mm}
 \includegraphics[width=1.06\textwidth]{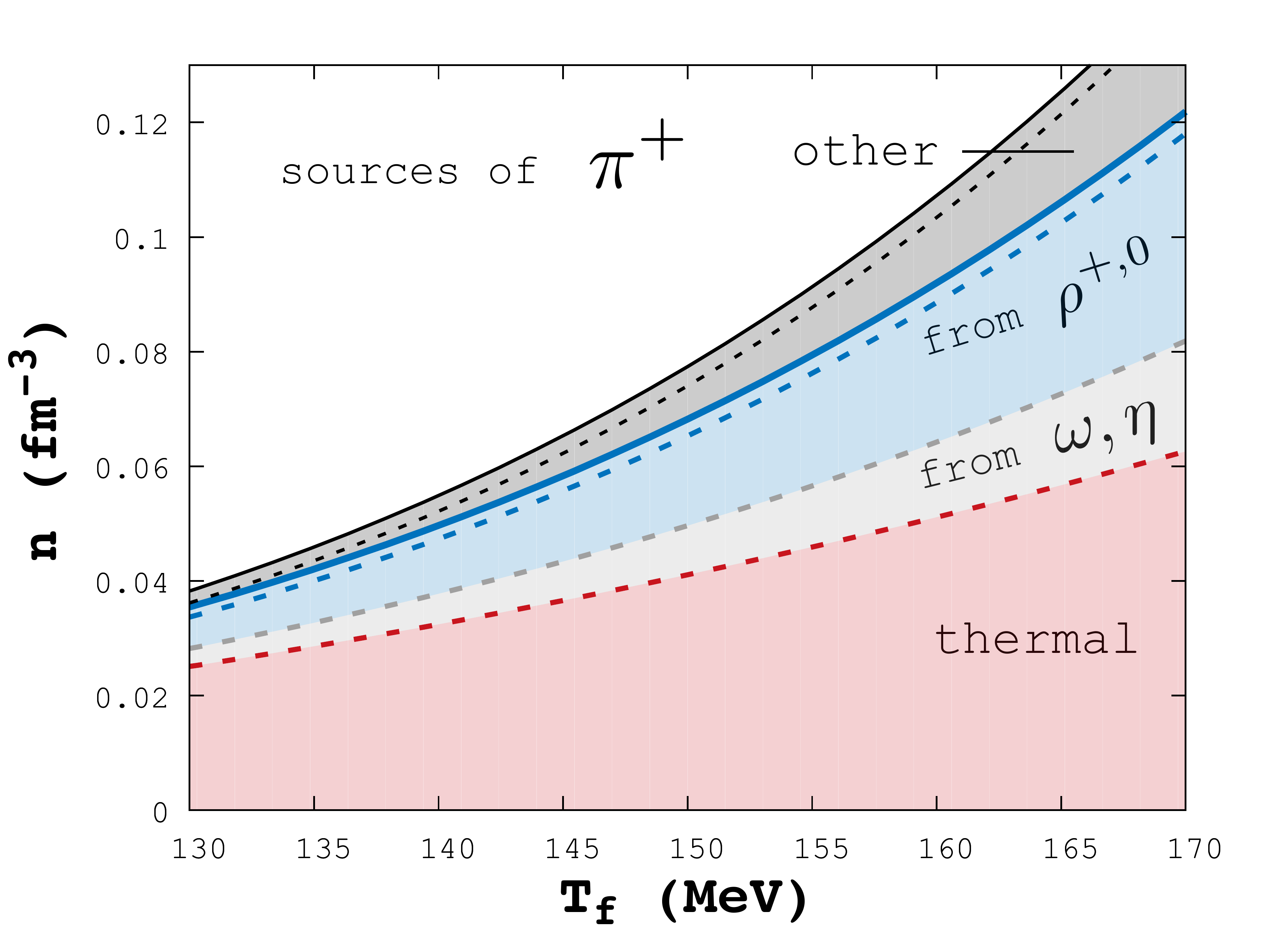}
\end{minipage}
\vspace*{-3mm}
	\caption{(colour online). { Left: $p_T$ spectra
  of $\pi^+$ originating from decays of $\rho$, the 
  $\pi K$(S- and P-wave) system, and the $\Delta(1232)$-channel of $\pi N$
  using both the S-matrix treatment and the zero width approximation at 
  $T = 155$ MeV temperature. The contribution from $\rho$ decays is calculated 
  also using the relativistic Breit-Wigner description of $\rho$'s. 
  Right: Contributions to pion density from various sources as function 
  of freeze-out temperature. In this calculation, the $\eta$
  and $\omega$ resonances have zero widths, and the S-matrix treatment
  has been applied to the system of $\rho$, and to the processes
  indicated as "other": the system of $\pi\pi$ (S-wave), $\pi K$
  (S- and P-wave) and the $\Delta(1232)$-channel of $\pi N$ (see
  text). In both figures, solid and dashed lines correspond to results
  of the S-matrix approach and the conventional zero-width approximation, 
  respectively.}}
\vspace{-0.3cm}
 \label{fig1}
\end{figure*}
%%%%%%%%%%%%%%%%%%%%% FIGURE %%%%%%%%%%%%%%%%%%%%%%%%%%%%%%%%

In the S-matrix formalism, the density of states per unit volume and
unit invariant mass $M$, assuming thermal equilibrium at temperature
$T$, is given
by~\cite{Beth:1937zz,Venugopalan:1992hy,Weinhold:1997ig,Broniowski:2003ax}
\begin{equation}
	\label{dos}
 \frac{\dif n_{IJ}}{\dif M}
 = \int \frac{\dif^3 p}{(2 \pi)^3}\, \frac{1}{2\pi}\, \mathcal{B}(M)\,
                                     f(E(M,p),T),
\end{equation}
where $f$ is the Bose-Einstein or Fermi-Dirac distribution, and
$\mathcal{B}(M)$ is an effective spectral weight,
\begin{equation}
	\label{B_func}
	\mathcal{B}(M) = 2 \, \frac{\dif \delta_{IJ}}{\dif M},
\end{equation}
derived from the scattering phase shift $\delta_{IJ}$, of the isospin
$I$ and spin $J$ channel.

In the elastic region ($M \lesssim 1 \, {\rm GeV}$), the empirical
phase shift~\cite{Protopopescu:1973sh,Estabrooks:1974vu,Froggatt:1977hu}
of the $(I=1, J=1)$ channel can be effectively described by a
phenomenological formula, inspired by a one-loop perturbative
calculation of the $\rho$ self-energy~\cite{Gale:1990pn,
  Herrmann:1993za},
\begin{equation}
	\label{fit_formula}
 \delta_{11}(M)
  =\tan^{-1}\left(- \frac{2}{3 M}\,\frac{\alpha_0}{1+c \, p_{\mathrm{CM}}^2} \,
                   \frac{p_{\mathrm{CM}}^3}{M^2-m_0^2}\right),
\end{equation}
where $p_{\mathrm{CM}}(M)=\frac{1}{2} \sqrt{M^2-4 m_\pi^2}$ is the
center-of-mass momentum of the scattering pions, and 
$\alpha_0 = 3.08$, $m_0 = 0.77 \, {\rm GeV}$, and 
$c = 0.59 \, {\rm GeV^{-2}}$ are the model parameters chosen to
reproduce not only the phase-shift data, but also the known value of
the P-wave scattering length. The phase shift and the scattering
length are related as
\begin{equation}
   a^1_1 = \frac{\delta_{11}}{p_{\mathrm{CM}}^3} \bigg|_{p_{\mathrm{CM}}\rightarrow0}.
 \end{equation}
We constrain the scattering length to $a^1_1 = 0.038 \, m_\pi^{-3}$,
matching the experimental value and chiral perturbation theory
prediction $a^1_1 = 0.038(2) \, m_\pi^{-3}$~\cite{Dumbrajs:1983jd} and
$0.037(10) \, m_\pi^{-3}$~\cite{Knecht:1995tr,Wang:2000ig},
respectively. This requirement is essential for the correct
description of the near-threshold behaviour of the density function,
introduced in Eq.~\eqref{B_func}.

An important feature of the current approach is the use of the
effective spectral weight $\mathcal{B}(M)$ instead of the standard
spectral function. This effective weight includes contributions from
both a pure $\rho$ state and the correlated $\pi \pi$ pair. The latter
tends to shift the strength of the weight function towards the low
invariant-mass region~\cite{Weinhold:1997ig}. Such a shift can
potentially translate into an enhancement of the low-$p_T$ daughter
pions from the decays of $\rho$ mesons.

To quantify this expectation, we evaluate the distribution of $\rho$'s
using the Cooper-Frye description~\cite{Cooper:1974mv}, with the
thermal distribution augmented by the effective spectral weight
$\mathcal{B}$ in Eq.~\eqref{B_func}, as
\begin{align}
\begin{aligned}
	\frac{\dif N_\rho}{\dif y\,p_T\,\dif p_T\, \dif\phi}
 & = \int {\dif M_\rho} \int \dif\sigma_\mu p_\rho^\mu  \,
          \frac{1}{2\pi}\,\mathcal{B}(M_\rho) \\
 & \times \frac{d_\rho}{(2 \pi)^3} \, f_\rho(p\cdot u,T),
\label{rho_spectrum}
\end{aligned}
\end{align}
where $f_\rho, d_\rho$ are respectively the Bose-Einstein distribution
and the spin degeneracy for $\rho$, and $u$ is the flow velocity. In
the case of a static source, the integration over the surface,
$\int\dif\sigma_\mu p^\mu$, becomes a simple multiplication by the
volume of the system, $V$, and by the energy of the particle, $E$. The
momentum spectrum of the decay pions can be evaluated by applying the
conventional decay
kinematics~\cite{Sollfrank:1991xm,Sollfrank:1990qz,Gorenstein:1987zm}
to the distribution of $\rho$'s from Eq.~(\ref{rho_spectrum}). For a
static source, one gets
{\begin{align}
\begin{aligned}
	\frac{\dif N_\pi^{de}}{\dif y\,p_T\,\dif p_T\, \dif\phi}
  & = V \int \dif{M_\rho}\, \frac{1}{2\pi}\, \mathcal{B}(M_\rho) \\
  &\hspace*{-6mm} \times \frac{M_\rho}{2 \, p_\pi p_{\mathrm{CM}}}
           \int_{E_\rho^{-}}^{E_\rho^{+}}\!\! \dif{E_\rho} \,E_\rho
           \frac{d_\rho}{(2 \pi)^3} \, f_\rho(E(M_\rho),T),
\label{pi_spectrum}
\end{aligned}
\end{align}}
where
\begin{align}
E_\rho^\pm = \frac{M_\rho}{2 \, m_\pi^2}(E_\pi M_\rho \pm 2 p_\pi p_{\mathrm{CM}}).
\end{align}
We evaluate the $p_T$ distributions at $T=155 \, {\rm MeV}$, in the
vicinity of the pseudocritical temperature obtained in the lattice
formulation of QCD~\cite{Borsanyi:2013bia,Bazavov:2014pvz}.

%%%%%%%%%%%%%%%%%%%%% FIGURE %%%%%%%%%%%%%%%%%%%%%%%%%%%%%%%%
\begin{figure*}[t]
\begin{minipage}{0.497\textwidth}
 \hspace*{-2mm}
 \includegraphics[width=1.06\textwidth]{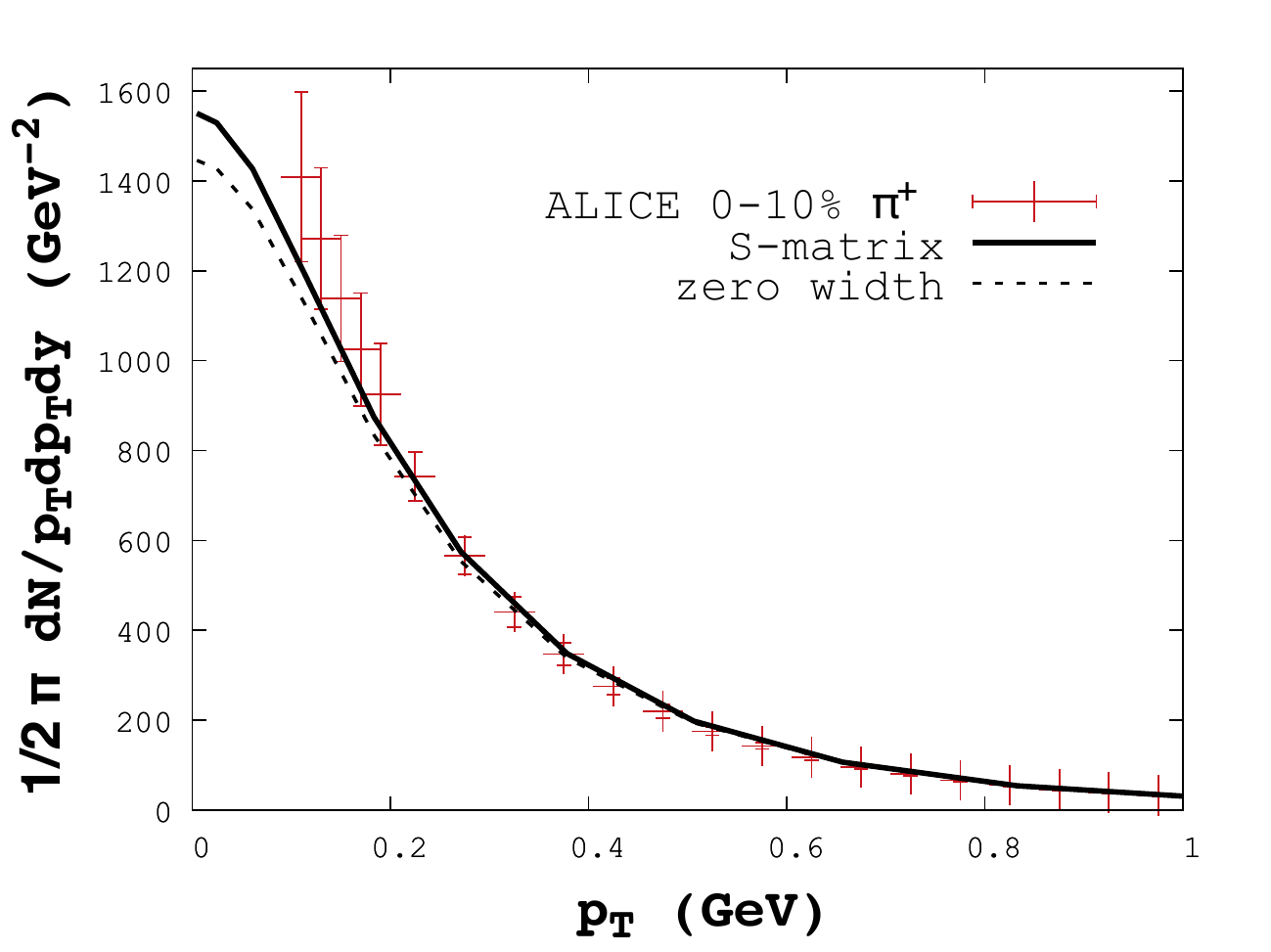}
\end{minipage}
 \hfill
\begin{minipage}{0.497\textwidth}
 \hspace*{-1mm}
 \includegraphics[width=1.06\textwidth]{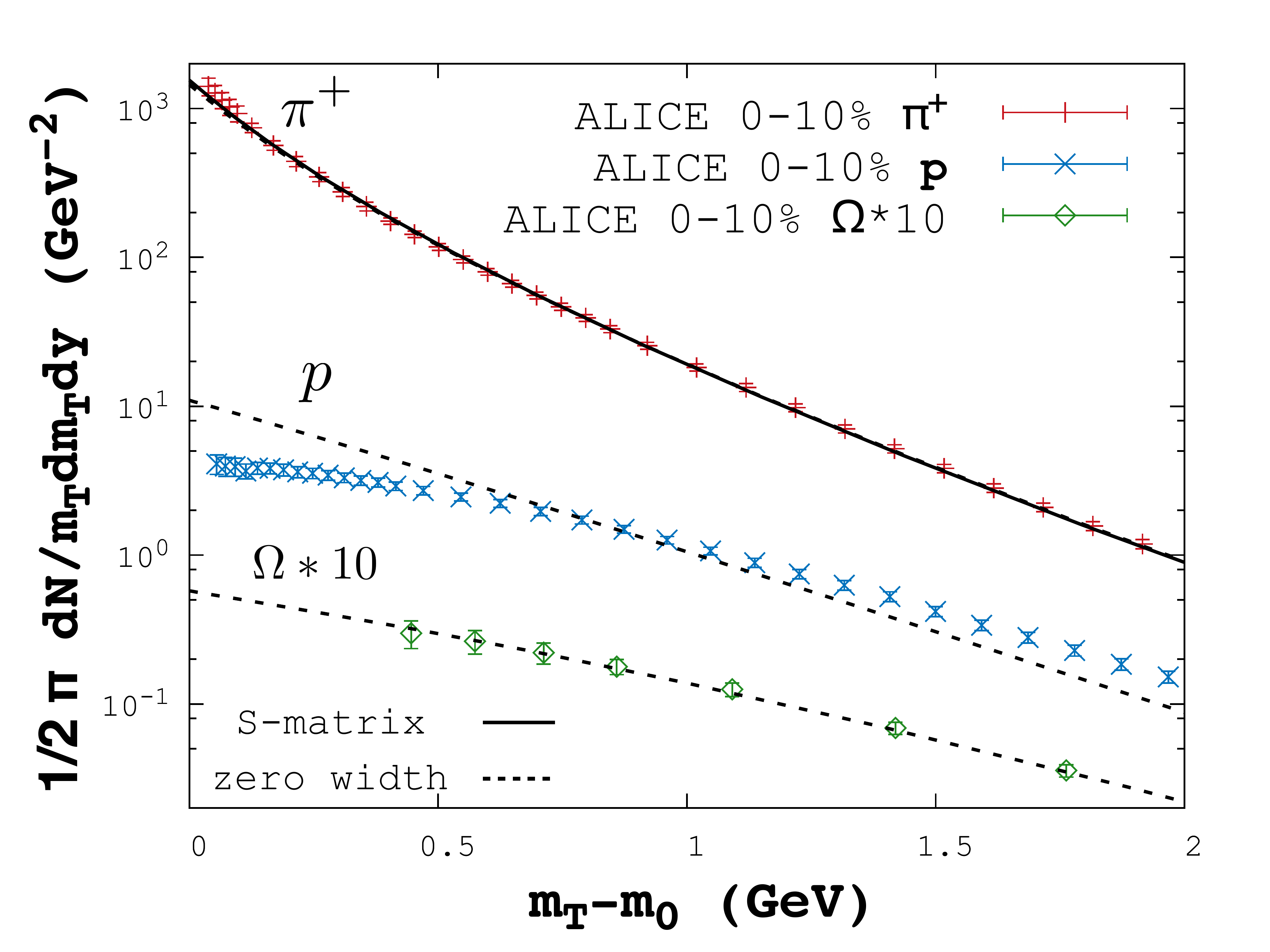}
\end{minipage}
\vspace*{-3mm}
\label{mTlog}
\caption{(colour online). \protect\small Left: The $p_T$ distribution
  of positive pions in 0-10\% most central $\sqrt{\sNN} = 2.76$ TeV
  Pb+Pb collisions as measured by the ALICE
  collaboration~\cite{Abelev:2013vea,ABELEV:2013zaa}, and fitted using
  a blast-wave model. Right: The \protect\small $m_T$ distribution of
  positive pions, protons and $\Omega$ baryons in 0-10\% most central
  $\sqrt{\sNN} = 2.76$ TeV Pb+Pb collisions as measured by the ALICE
  collaboration~\cite{Abelev:2013vea,ABELEV:2013zaa}, and fitted using
  a blast-wave model. In both panels the solid lines correspond to the
  S-matrix approach result and the dashed lines to the conventional
  zero-width approximation.}
\label{fig2}
\end{figure*}
%%%%%%%%%%%%%%%%%%%%% FIGURE %%%%%%%%%%%%%%%%%%%%%%%%%%%%%%%%

{ In Fig.~\ref{fig1}-Left we show the rapidity and
  azimuthal angle integrated transverse momentum spectra of $\pi^+$
  originating from $\rho$ decays. The $\rho$'s are treated as zero-width
  particles, particles with the standard Breit-Wigner width, or
  according to the S-matrix approach introduced in
  Eq.~(\ref{pi_spectrum}). The latter description leads to a
  substantial enhancement of the pion decay spectra.}
The effect is most prominent in the low-$p_T$ region of the decay
pions, where at $p_T \approx 0$ one observes a factor of two increase
of the differential pion yield. Note that at larger values of the
transverse momentum the spectrum of decay pions is practically
unaffected by the width of $\rho$.
{ For future reference, we also present results on decay $\pi^+$
  spectra from the system of $\pi K$ interaction (sum of S- and
  P-wave) and from $\pi N$ interaction in the $\Delta$-channel. In
  all the channels studied we find overall enhancement of low-$p_T$
  pions in the S-matrix approach compared to the zero-width
  results. Nevertheless, the difference is most noticeable in the
  $\rho$ sector.}

%further justifying the importance of $\rho$'s.

{ 
%In the HRG, the $\rho$ mesons are not the only resonances which
%contribute to the overall pion production. 
To illustrate the effect of
leading resonances on the pion yield in the HRG, we show in Fig.~\ref{fig1}-right
the temperature dependence of the contributions from various sources
to pion density after resonance decays.  For this analysis, we have
included the three-body decays of zero-width $\eta$ and $\omega$ (with
branching ratios of $0.228$ and $0.893$, respectively). Furthermore,
we have applied the S-matrix treatment to the $\pi\pi$ (S-wave) and
$\pi K$ (S- and P-wave) systems and the $\Delta(1232)$-channel of 
$\pi N$. At $T = 155 \, {\rm MeV}$, when no heavier resonances are
included, the relative abundance of $\pi^+$ from $\rho$ decay is
$ 25.1 \%$, while the thermal pion yield remains dominant at
$49.4 \%$. Three-body decays considered constitute $12.2 \%$, 
and the sum of the rest of two-body channels we treated give $13.3 \%$
of the total yield. The S-matrix treatment significantly affects the yield of pions from
$\rho$ decays, resulting in its increase by approximately $15\%$, whereas
the effect is smaller for other channels considered. However, because of the contribution from all the other sources,
the overall change in the final pion yield due to the S-matrix
approach is only a few per cent.}

In general, on the level of particle yields, and at higher
temperatures $T> 100$ MeV, the zero-width treatment of resonances
gives comparable results to the S-matrix
approach~\cite{Venugopalan:1992hy} despite the fact that the phase
shifts in most cases do not resemble a step function and the
assumption of a zero (and at times even a narrow) width is strictly
speaking not justified. However, as already seen in Fig. 1, essential
differences can appear when $p_T$-differential observables of
individual resonance channels are studied. Evidently, the more
physical treatment by the S-matrix formulation is needed for precision
calculations of particle spectra, as e.g. in modeling data in
heavy-ion collisions.

In a realistic heavy-ion collision, however, the situation is further
complicated by the expansion of the system, and the presence of all
the other resonances. To gauge whether the S-matrix description of
$\rho$ mesons would affect the pion distributions observed in
heavy-ion collisions, we describe the system using a blast-wave
model~\cite{Schnedermann:1993ws}. There, the thermal source is
assumed to be a boost-invariant~\cite{Bjorken:1982qr} cylindrically
symmetric transversely expanding tube of radius $R$, from which
particles are emitted at constant longitudinal proper time $\tau$ with
the radial flow velocity $v(r)=v_{max}(r/R)$.

We calculate the distributions of all the resonances in the Particle
Data Book up to the 2 GeV mass, apply the two- and three-body decay
kinematics, and sum the contributions to the spectrum of thermal
pions. We take advantage of the recent finding in the dynamical model
calculations in heavy-ion collisions that the pion $p_T$-distribution
changes only very little during the subsequent evolution in the
hadronic phase~\cite{Xu:2016hmp}. Thus, we fix the freeze-out
temperature at $T=155$ MeV, which coincides with the chiral crossover
in LQCD. The further parameters of the blast-wave model, $\tau = 13.7$
fm, $R= 10$ fm, and $v_\mathrm{max} = 0.8$ were chosen to get the best
description of spectra for positive pions in 0-10\% most central
$\sqrt{\sNN} = 2.76$ TeV Pb+Pb collisions as measured by the ALICE
collaboration. The above freeze-out temperature and the resulting
volume of the fireball, $V\simeq 4300 $ fm$^3$, are consistent with
that obtained previously in the HRG model description of hadron
production yields and some fluctuation observables in heavy-ion
collisions at the LHC \cite{f2,k3}.

The resulting pion distribution is shown in the left panel of
Fig.~\ref{fig2}. In this calculation the conventional zero-width
treatment of $\rho$'s leads to a distribution which underestimates
the data in the low-$p_T$ region ($p_T \lton 200\,\mathrm{MeV}$). When
$\rho$ mesons are treated according to the S-matrix description, there
is a clear, up to 7$\%$, increase of the low-$p_T$ pions, which is
sufficient to reach the data.

To check further the quality of the model parametrisation, we also
show in the right panel of Fig.~\ref{fig2} the pion, proton and
$\Omega$ baryon distributions in a broader $m_T$-range.  As seen in
this figure, the pion data are well described up to $m_T\simeq 2$ GeV,
and the model predictions are also consistent with the data for the
$\Omega$ distribution.  These results verify the chosen values for
temperature and volume, and they are also consistent with the idea
that $\Omega$ baryons hardly rescatter in the hadronic
phase~\cite{nu,Takeuchi:2015ana}, and thus their spectra are fixed at
the phase boundary~\cite{nu}.  On the other hand, the proton
distribution is steeper than the data, and the overall yield of
protons is larger than the experimental value. The observed deviation
on the level of proton yield is already discussed in the
literature~\cite{f2}. The deviations in the proton spectrum could be
possibly due to their further rescattering during evolution in the
hadronic phase~\cite{Takeuchi:2015ana,Hirano:2005wx}.

In conclusion, we have investigated how the explicit treatment of the
$\rho$-meson width affects the pion yield and $p_T$ distribution in
$\sqrt{\sNN} = 2.76$ TeV Pb+Pb collisions at LHC. We have used the
S-matrix approach to describe $\rho$ mesons, and found that compared
to the conventional zero-width treatment the pion yield increases,
particularly at low values of transverse momentum. This indicates that
the observed enhancement of low-$p_T$ pions may be possibly explained
in fluid-dynamical calculations by a proper implementation of the
width of resonances within the S-matrix approach. However, the S-matrix
treatment of $\rho$'s alone may not be fully sufficient.

A natural extension of this work is to apply a more complete model for
the fluid dynamical calculations~\cite{Heinz:2013th,Gale:2013da,
  Hirano:2012kj,deSouza:2015ena,Jaiswal:2016hex}, as well as, to
account for a possible medium modification on the phase
shifts. Essential in-medium effects for $\rho$ mesons are suggested by
studies based on many-body Green's function~\cite{Herrmann:1993za,
  Rapp:1995zy,Rapp:1997fs,Klingl:1997kf,Leupold:1997dg,Rapp:2009yu}. 
This, together with the S-matrix treatment of three-body decays, can
presumably further increase the pion yields in the low-$p_T$
region. We leave this as a matter of future investigation.
Nevertheless, even in their present level, our results demonstrate the
importance of the proper treatment of resonances in modeling heavy-ion
collisions, and the need to improve on the customary hadron resonance
gas models for precision calculations of particle spectra at low
values of transverse momentum.
{
%These studies are also important in  hydrodynamics-hybrid models for
%particle production in heavy ion collisions  when describing
%particalization  of a quark-gluon plasma as an input to hadronic
%transport.
%
These studies are also important in hydrodynamics-cascade hybrid 
models~\cite{Hirano:2012kj,Huovinen:2012is}
for particle production in heavy ion collisions when describing
particalization of the fluid as an input to hadronic transport.
}

We acknowledge fruitful discussions with Bengt Friman. K.R.~also
acknowledges interesting discussions with Peter Braun-Munzinger,
Stefen Bass and members of the Nuclear Physics Group at Duke
University. P.M.L.~acknowledges the support of the Extreme Matter
Institute (EMMI). K.M.~was supported by the Grants-in-Aid for
Scientific Research on Innovative Area from MEXT (No.~24105008) and
Grants-in-Aid for Scientific Research from JSPS (No.~16K05349). This
work was supported by the National Science Center, Poland under
grants: Maestro DEC-2013/10/A/ST2/00106 and Polonez
DEC-2015/19/P/ST2/03333, as well as has received funding from the
European Union's Horizon 2020 research and innovation programme under
the Marie Sk\l odowska-Curie grant agreement No 665778.


\vspace*{-0.5cm}
\begin{thebibliography}{20}


\bibitem{Abelev:2013vea}
  B.~Abelev {\it et al.} [ALICE Collaboration],
  %``Centrality dependence of $\pi$, K, p production in Pb-Pb collisions at
  % $\sqrt{s_{NN}}$ = 2.76 TeV,''
  Phys.\ Rev.\ C {\bf 88} (2013) 044910.
  %%CITATION = doi:10.1103/PhysRevC.88.044910;%%

\bibitem{Begun:2013nga}
  V.~Begun, W.~Florkowski and M.~Rybczynski,
  %``Explanation of hadron transverse-momentum spectra in heavy-ion collisions
  % at $\sqrt s_{NN} =$ 2.76 TeV within chemical non-equilibrium statistical
  % hadronization model,''
  Phys.\ Rev.\ C {\bf 90} (2014) 014906
  [arXiv:1312.1487 [nucl-th]].
  %%CITATION = doi:10.1103/PhysRevC.90.014906;%%

\bibitem{Molnar:2014zha}
  E.~Molnar, H.~Holopainen, P.~Huovinen and H.~Niemi,
  %``Influence of temperature-dependent shear viscosity on elliptic
  %flow at backward and forward rapidities in ultrarelativistic
  %heavy-ion collisions,''
  Phys.\ Rev.\ C {\bf 90} (2014) 044904
  [arXiv:1407.8152 [nucl-th]].
  %%CITATION = doi:10.1103/PhysRevC.90.044904;%%

\bibitem{Sollfrank:1991xm}
  J.~Sollfrank, P.~Koch and U.~W.~Heinz,
  %``Is there a low p(T) 'anomaly' in the pion momentum spectra from
  % relativistic nuclear collisions?,''
  Z.\ Phys.\ C {\bf 52} (1991) 593.
  %%CITATION = doi:10.1007/BF01562334;%%

\bibitem{Venugopalan:1992hy}
  R.~Venugopalan and M.~Prakash,
  %``Thermal properties of interacting hadrons,''
  Nucl.\ Phys.\ A {\bf 546} (1992) 718.
  %%CITATION = doi:10.1016/0375-9474(92)90005-5;%%

\bibitem{Gerber:1988tt}
  P.~Gerber and H.~Leutwyler,
  %``Hadrons Below the Chiral Phase Transition,''
  Nucl.\ Phys.\ B {\bf 321} (1989) 387.
  %%CITATION = doi:10.1016/0550-3213(89)90349-0;%%

\bibitem{Weinhold:1997ig}
  W.~Weinhold, B.~Friman and W.~N\"orenberg,
  %``Thermodynamics of Delta resonances,''
  Phys.\ Lett.\ B {\bf 433} (1998) 236
  [nucl-th/9710014].
  %%CITATION = NUCL-TH/9710014;%%

\bibitem{n1}
  D.~H.~Rischke,
  %``Remarks on the extraction of freezeout parameters,''
  Nucl.\ Phys.\ A {\bf 698} (2002) 153.
  %%CITATION = doi:10.1016/S0375-9474(01)01359-8;%%

\bibitem{f1}
  P.~Braun-Munzinger, K.~Redlich, and  J.~Stachel, 
  Particle production in heavy ion collisions, in:
  R.~C.~Hwa and X.-N.~Wang (Eds.), Quark-Gluon Plasma 3,
  World Scientific, Singapore, 2004, pp.~491-599.
  e-Print: nucl-th/0304013.
  %%CITATION = NUCL-TH/0304013;%%

\bibitem{f2}
  J.~Stachel, A.~Andronic, P.~Braun-Munzinger and K.~Redlich,
  %``Confronting LHC data with the statistical hadronization model,''
  J.\ Phys.\ Conf.\ Ser.\  {\bf 509} (2014) 012019.
  %%CITATION = doi:10.1088/1742-6596/509/1/012019;%%

\bibitem{f3}
  M.~Floris,
  %``Hadron yields and the phase diagram of strongly interacting matter,''
  Nucl.\ Phys.\ A {\bf 931} (2014) 103.
  %%CITATION = doi:10.1016/j.nuclphysa.2014.09.002;%%

\bibitem{f4}
  A.~Andronic,
  %``An overview of the experimental study of quark-gluon matter in
  %high-energy nucleus-nucleus collisions,''
  Int.\ J.\ Mod.\ Phys.\ A {\bf 29} (2014) 1430047.
  %%CITATION = doi:10.1142/S0217751X14300476;%%



%%%%%%%%%%%%%%%%%%%%%%%%%%%%%%%%%%%%%%%%%%%%%%%%%%
\bibitem{cleymans}
 I.~Kraus, J.~Cleymans, H.~Oeschler and K.~Redlich,
 %``Particle production in p-p collisions and prediction for LHC energy,''
 Phys.\ Rev.\ C {\bf 79}, 014901 (2009)
 [arXiv:0808.0611 [hep-ph]].
 %%CITATION = doi:10.1103/PhysRevC.79.014901;%%

%\cite{Becattini:1997rv}
\bibitem{Becattini:1997rv} 
  F.~Becattini and U.~W.~Heinz,
  %``Thermal hadron production in p p and p anti-p collisions,''
  Z.\ Phys.\ C {\bf 76}, 269 (1997)
  Erratum: [Z.\ Phys.\ C {\bf 76}, 578 (1997)]
  %doi:10.1007/s002880050551
  [hep-ph/9702274].
  %%CITATION = doi:10.1007/s002880050551;%%
  %326 citations counted in INSPIRE as of 22 Mar 2017


%\cite{Becattini:1997ii}
\bibitem{Becattini:1997ii} 
  F.~Becattini, M.~Gazdzicki and J.~Sollfrank,
  %``On chemical equilibrium in nuclear collisions,''
  Eur.\ Phys.\ J.\ C {\bf 5}, 143 (1998)
  %doi:10.1007/s100529800831, 10.1007/s100520050256
  [hep-ph/9710529].
  %%CITATION = doi:10.1007/s100529800831, 10.1007/s100520050256;%%
  %227 citations counted in INSPIRE as of 22 Mar 2017



%\cite{Kisiel:2005hn}
\bibitem{Kisiel:2005hn} 
  A.~Kisiel, T.~Taluc, W.~Broniowski and W.~Florkowski,
  %``THERMINATOR: THERMal heavy-IoN generATOR,''
  Comput.\ Phys.\ Commun.\  {\bf 174}, 669 (2006)
  %doi:10.1016/j.cpc.2005.11.010
  [nucl-th/0504047].
  %%CITATION = doi:10.1016/j.cpc.2005.11.010;%%
  %179 citations counted in INSPIRE as of 22 Mar 2017


%\cite{Chojnacki:2011hb}
\bibitem{Chojnacki:2011hb} 
  M.~Chojnacki, A.~Kisiel, W.~Florkowski and W.~Broniowski,
  %``THERMINATOR 2: THERMal heavy IoN generATOR 2,''
  Comput.\ Phys.\ Commun.\  {\bf 183}, 746 (2012).
  %doi:10.1016/j.cpc.2011.11.018
  %[arXiv:1102.0273 [nucl-th]].
  %%CITATION = doi:10.1016/j.cpc.2011.11.018;%%
  %90 citations counted in INSPIRE as of 22 Mar 2017





%%%%%%%%%%%%%%%%%%%%%%%%%%%%%%%%%%%%%%%%%%%%%%%%%%


\bibitem{Karsch:2003vd}
  F.~Karsch, K.~Redlich and A.~Tawfik,
  %``Hadron resonance mass spectrum and lattice QCD thermodynamics,''
  Eur.\ Phys.\ J.\ C {\bf 29} (2003) 549
  [hep-ph/0303108].
  %%CITATION = doi:10.1140/epjc/s2003-01228-y;%%

\bibitem{Huovinen:2009yb}
  P.~Huovinen and P.~Petreczky,
  %``QCD Equation of State and Hadron Resonance Gas,''
  Nucl.\ Phys.\ A {\bf 837} (2010) 26
  [arXiv:0912.2541 [hep-ph]].
  %%CITATION = doi:10.1016/j.nuclphysa.2010.02.015;%%

\bibitem{k1}
  A.~Bazavov {\it et al.},
  %``Freeze-out Conditions in Heavy Ion Collisions from QCD Thermodynamics,''
  Phys.\ Rev.\ Lett.\ {\bf 109} (2012) 192302.
  %%CITATION = doi:10.1103/PhysRevLett.109.192302;%%

\bibitem{k2}
  C.~Ratti {\it et al.} [Wuppertal-Budapest Collaboration],
  %``Recent results on QCD thermodynamics: lattice QCD versus Hadron
  %Resonance Gas model,''
  Nucl.\ Phys.\ A {\bf 855} (2011) 253.
  %%CITATION = doi:10.1016/j.nuclphysa.2011.02.052;%%

\bibitem{k3}
P.~Braun-Munzinger, A.~Kalweit, K.~Redlich and J.~Stachel,
  %``Confronting fluctuations of conserved charges in central nuclear
  %collisions at the LHC with predictions from Lattice QCD,''
  Phys.\ Lett.\ B {\bf 747} (2015) 292.
  %%CITATION = doi:10.1016/j.nuclphysa.2010.02.015;%%

\bibitem{Broniowski:2015oha}
  W.~Broniowski, F.~Giacosa and V.~Begun,
  %``Cancellation of the $\sigma$ meson in thermal models,''
  Phys.\ Rev.\ C {\bf 92} (2015) 034905
  [arXiv:1506.01260 [nucl-th]].
  %%CITATION = doi:10.1103/PhysRevC.92.034905;%%

\bibitem{Friman:2015zua}
  B.~Friman, P.~M.~Lo, M.~Marczenko, K.~Redlich and C.~Sasaki,
  %``Strangeness fluctuations from $K-\pi$ interactions,''
  Phys.\ Rev.\ D {\bf 92} (2015) 074003
  [arXiv:1507.04183 [hep-ph]].
  %%CITATION = doi:10.1103/PhysRevD.92.074003;%%

\bibitem{Dashen:1969ep}
  R.~Dashen, S.~K.~Ma and H.~J.~Bernstein,
  %``S Matrix formulation of statistical mechanics,''
  Phys.\ Rev.\ {\bf 187} (1969) 345.
  %%CITATION = doi:10.1103/PhysRev.187.345;%%

\bibitem{Beth:1937zz}
  E.~Beth and G.~Uhlenbeck,
  %``The quantum theory of the non-ideal gas. II. Behaviour at low
  %temperatures,''
  Physica {\bf 4} (1937) 915.
  %%CITATION = PHYSA,4,915;%%

\bibitem{Denisenko:1986bb}
  K.~G.~Denisenko and S.~Mrowczynski,
  %``Deltas in Hadron Gas,''
  Phys.\ Rev.\ C {\bf 35} (1987) 1932.
  %%CITATION = doi:10.1103/PhysRevC.35.1932;%%

\bibitem{Broniowski:2003ax}
  W.~Broniowski, W.~Florkowski and B.~Hiller,
  %``Thermal analysis of production of resonances in relativistic
  % heavy ion collisions,''
  Phys.\ Rev.\ C {\bf 68} (2003) 034911
  [nucl-th/0306034].
  %%CITATION = doi:10.1103/PhysRevC.68.034911;%%

\bibitem{Protopopescu:1973sh}
  S.~D.~Protopopescu {\it et al.},
  %``Pi pi Partial Wave Analysis from Reactions pi+ p ---> pi+ pi-
  %Delta++ and pi+ p ---> K+ K- Delta++ at 7.1-GeV/c,''
  Phys.\ Rev.\ D {\bf 7} (1973) 1279.
  %%CITATION = doi:10.1103/PhysRevD.7.1279;%%

\bibitem{Estabrooks:1974vu}
  P.~Estabrooks and A.~D.~Martin,
  %``pi pi Phase Shift Analysis Below the K anti-K Threshold,''
  Nucl.\ Phys.\ B {\bf 79} (1974) 301.
  %%CITATION = doi:10.1016/0550-3213(74)90488-X;%%

\bibitem{Froggatt:1977hu}
  C.~D.~Froggatt and J.~L.~Petersen,
  %``Phase Shift Analysis of pi+ pi- Scattering Between 1.0-GeV and
  %1.8-GeV Based on Fixed Momentum Transfer Analyticity. 2.,''
  Nucl.\ Phys.\ B {\bf 129} (1977) 89.
  %%CITATION = doi:10.1016/0550-3213(77)90021-9;%%

\bibitem{Gale:1990pn}
  C.~Gale and J.~I.~Kapusta,
  %``Vector dominance model at finite temperature,''
  Nucl.\ Phys.\ B {\bf 357} (1991) 65.
  %%CITATION = doi:10.1016/0550-3213(91)90459-B;%%

\bibitem{Herrmann:1993za}
  M.~Herrmann, B.~L.~Friman and W.~Norenberg,
  %``Properties of rho mesons in nuclear matter,''
  Nucl.\ Phys.\ A {\bf 560} (1993) 411.
  %%CITATION = doi:10.1016/0375-9474(93)90105-7;%%

\bibitem{Dumbrajs:1983jd}
  O.~Dumbrajs {\it et al.},
  %``Compilation of Coupling Constants and Low-Energy Parameters. 1982
  %Edition,''
  Nucl.\ Phys.\ B {\bf 216} (1983) 277.
  %%CITATION = doi:10.1016/0550-3213(83)90288-2;%%

\bibitem{Knecht:1995tr}
  M.~Knecht, B.~Moussallam, J.~Stern and N.~H.~Fuchs,
  %``The Low-energy pi pi amplitude to one and two loops,''
  Nucl.\ Phys.\ B {\bf 457} (1995) 513
  [hep-ph/9507319].
  %%CITATION = doi:10.1016/0550-3213(95)00515-3;%%

 \bibitem{Wang:2000ig}
  X.~J.~Wang and M.~L.~Yan,
  %``Rigorous effective field theory study on pion form-factor,''
  hep-ph/0005184.
  %%CITATION = HEP-PH/0005184;%%

\bibitem{Cooper:1974mv}
  F.~Cooper, G.~Frye,
  % ``Comment On The Single Particle Distribution In The Hydrodynamic And
  % Statistical Thermodynamic Models Of Multiparticle Production,''
  Phys.\ Rev.\ D {\bf 10} (1974) 186.
  %%CITATION = PHRVA,D10,186;%%

\bibitem{Sollfrank:1990qz}
  J.~Sollfrank, P.~Koch and U.~W.~Heinz,
  %``The Influence of resonance decays on the P(t) spectra from heavy
  %ion collisions,''
  Phys.\ Lett.\ B {\bf 252} (1990) 256.
  %%CITATION = doi:10.1016/0370-2693(90)90870-C;%%

\bibitem{Gorenstein:1987zm}
  M.~I.~Gorenstein, M.~S.~Tsai and S.~N.~Yang,
  %``Freezeout conditions and pion spectrum in heavy ion collisions,''
  Phys.\ Rev.\ C {\bf 51} (1995) 1465.
 % doi:10.1103/PhysRevC.51.1465
  %%CITATION = doi:10.1103/PhysRevC.51.1465;%%

\bibitem{Borsanyi:2013bia}
  S.~Borsanyi, Z.~Fodor, C.~Hoelbling, S.~D.~Katz, S.~Krieg and K.~K.~Szabo,
  %``Full result for the QCD equation of state with 2+1 flavors,''
  Phys.\ Lett.\ B {\bf 730} (2014) 99
  [arXiv:1309.5258 [hep-lat]].
  %%CITATION = doi:10.1016/j.physletb.2014.01.007;%%

\bibitem{Bazavov:2014pvz}
  A.~Bazavov {\it et al.} [HotQCD Collaboration],
  %``Equation of state in ( 2+1 )-flavor QCD,''
  Phys.\ Rev.\ D {\bf 90} (2014) 094503
  [arXiv:1407.6387 [hep-lat]].
  %%CITATION = doi:10.1103/PhysRevD.90.094503;%%

\bibitem{Schnedermann:1993ws}
  E.~Schnedermann, J.~Sollfrank and U.~W.~Heinz,
  %``Thermal phenomenology of hadrons from 200-A/GeV S+S collisions,''
  Phys.\ Rev.\ C {\bf 48} (1993) 2462
  [nucl-th/9307020].
  %%CITATION = doi:10.1103/PhysRevC.48.2462;%%

\bibitem{Bjorken:1982qr}
  J.~D.~Bjorken,
  % ``Highly Relativistic Nucleus-Nucleus Collisions: The Central Rapidity
  % Region,''
  Phys.\ Rev.\  D {\bf 27} (1983) 140.
  %%CITATION = PHRVA,D27,140;%%

\bibitem{Xu:2016hmp}
  H.~j.~Xu, Z.~Li and H.~Song,
  %``High-order flow harmonics of identified hadrons in 2.76A TeV Pb +
  %Pb collisions,''
  Phys.\ Rev.\ C {\bf 93} (2016) 064905
  [arXiv:1602.02029 [nucl-th]].
  %%CITATION = doi:10.1103/PhysRevC.93.064905;%%

\bibitem{ABELEV:2013zaa}
  B.~B.~Abelev {\it et al.} [ALICE Collaboration],
  %``Multi-strange baryon production at mid-rapidity in Pb-Pb
  %collisions at $\sqrt{s_{NN}}$ = 2.76 TeV,''
  Phys.\ Lett.\ B {\bf 728} (2014) 216
  Erratum: [Phys.\ Lett.\ B {\bf 734} (2014) 409]
  [arXiv:1307.5543 [nucl-ex]].
  %%CITATION =
  %%doi:10.1016/j.physletb.2014.05.052,10.1016/j.physletb.2013.11.048;%%

\bibitem{Takeuchi:2015ana}
  S.~Takeuchi, K.~Murase, T.~Hirano, P.~Huovinen and Y.~Nara,
  %``Effects of hadronic rescattering on multistrange hadrons in
  %high-energy nuclear collisions,''
  Phys.\ Rev.\ C {\bf 92} (2015) 044907
  [arXiv:1505.05961 [nucl-th]].
  %%CITATION = doi:10.1103/PhysRevC.92.044907;%%

\bibitem{nu}
  N.~Xu and Z.~B.~Xu,
  %``Transverse dynamics at RHIC,''
  Nucl.\ Phys.\ A {\bf 715} (2003) 587.
  %%CITATION = doi:10.1016/S0375-9474(02)01538-5;%%

\bibitem{Hirano:2005wx}
  T.~Hirano and M.~Gyulassy,
  %``Perfect fluidity of the quark gluon plasma core as seen through
  %its dissipative hadronic corona,''
  Nucl.\ Phys.\ A {\bf 769} (2006) 71
  [nucl-th/0506049].
  %%CITATION = doi:10.1016/j.nuclphysa.2006.02.005;%%

\bibitem{Heinz:2013th} 
  U.~W.Heinz and R.~Snellings,
  %``Collective flow and viscosity in relativistic heavy-ion collisions,''
  Annu.\ Rev.\ Nucl.\ Part.\ Sci.\ {\bf 63} (2013) 123
  [arXiv:1301.2826 [nucl-th]].
  %%CITATION = ARXIV:1301.2826;%%

\bibitem{Gale:2013da} 
  C.~Gale, S.~Jeon and B.~Schenke, 
  %``Hydrodynamic Modeling of Heavy-Ion Collisions,''
  Int.\ J.\ Mod.\ Phys.\ A \textbf{28} (2013) 1340011
  [arXiv:1301.5893 [nucl-th]].
  %%CITATION = ARXIV:1301.5893;%%

\bibitem{Hirano:2012kj}
  T.~Hirano, P.~Huovinen, K.~Murase and Y.~Nara,
  %``Integrated Dynamical Approach to Relativistic Heavy Ion Collisions,''
  Prog.\ Part.\ Nucl.\ Phys.\  {\bf 70} (2013) 108
  [arXiv:1204.5814 [nucl-th]].
  %%CITATION = doi:10.1016/j.ppnp.2013.02.002;%%

\bibitem{deSouza:2015ena}
  R.~Derradi de Souza, T.~Koide and T.~Kodama,
  %``Hydrodynamic Approaches in Relativistic Heavy Ion Reactions,''
  Prog.\ Part.\ Nucl.\ Phys.\  {\bf 86} (2016) 35
  [arXiv:1506.03863 [nucl-th]].
  %%CITATION = doi:10.1016/j.ppnp.2015.09.002;%%

\bibitem{Jaiswal:2016hex}
  A.~Jaiswal and V.~Roy,
  %``Relativistic hydrodynamics in heavy-ion collisions: general
  %aspects and recent developments,''
  arXiv:1605.08694 [nucl-th].
  %%CITATION = ARXIV:1605.08694;%%

\bibitem{Rapp:1995zy}
  R.~Rapp, G.~Chanfray and J.~Wambach,
  %``Medium modifications of the rho meson at CERN SPS energies,''
  Phys.\ Rev.\ Lett.\  {\bf 76} (1996) 368
  [hep-ph/9508353].
  %%CITATION = doi:10.1103/PhysRevLett.76.368;%%

\bibitem{Rapp:1997fs}
  R.~Rapp, G.~Chanfray and J.~Wambach,
  %``Rho meson propagation and dilepton enhancement in hot hadronic matter,''
  Nucl.\ Phys.\ A {\bf 617} (1997) 472
  [hep-ph/9702210].
  %%CITATION = doi:10.1016/S0375-9474(97)00137-1;%%

\bibitem{Klingl:1997kf}
  F.~Klingl, N.~Kaiser and W.~Weise,
  %``Current correlation functions, QCD sum rules and vector mesons in
  %baryonic matter,''
  Nucl.\ Phys.\ A {\bf 624} (1997) 527
  [hep-ph/9704398].
  %%CITATION = doi:10.1016/S0375-9474(97)88960-9;%%

\bibitem{Leupold:1997dg}
  S.~Leupold, W.~Peters and U.~Mosel,
  %``What QCD sum rules tell about the rho meson,''
  Nucl.\ Phys.\ A {\bf 628} (1998) 311
  [nucl-th/9708016].
  %%CITATION = doi:10.1016/S0375-9474(97)00634-9;%%

\bibitem{Rapp:2009yu}
  R.~Rapp, J.~Wambach and H.~van Hees,
  %``The Chiral Restoration Transition of QCD and Low Mass Dileptons,''
  Landolt-Bornstein {\bf 23} (2010) 134
  [arXiv:0901.3289 [hep-ph]].
  %%CITATION = doi:10.1007/978-3-642-01539-7_6;%%. 

\bibitem{Huovinen:2012is}
  P.~Huovinen and H.~Petersen,
  %``Particlization in hybrid models,''
  Eur.\ Phys.\ J.\ A {\bf 48} (2012) 171
  [arXiv:1206.3371 [nucl-th]].
  %%CITATION = doi:10.1140/epja/i2012-12171-9;%%

\end{thebibliography}
\end{document}